\documentstyle[12pt,epsf]{article}

\begin{document}

\begin{center}

{\Large \bf From spinning to non-spinning cosmic string spacetime} \\

\vspace{2cm}

V. A. De Lorenci${}^{*}$

\vspace{0.41cm}
{\it Instituto de Ci\^encias --
Escola Federal de Engenharia de Itajub\'a \\
Av. BPS 1303 Pinheirinho, 37500-000 Itajub\'a, MG -- Brazil} 

\vspace{0.81cm}
R. D. M. De Paola${}^{\dag}$ and 
N. F. Svaiter ${}^{\ddag}$ \\

\vspace{0.41cm} 
{\it Centro Brasileiro de Pesquisas F\'{\i}sicas\\
Rua Dr. Xavier Sigaud 150, Urca,
22290-180 Rio de Janeiro, RJ -- Brazil} \\

\vspace{2cm}

\begin{abstract}

We analyze the properties of a fluid generating a spinning
cosmic string spacetime with flat limiting cases corresponding to
a constant angular momentum in the infinite past and static 
configuration in the infinite future.
The spontaneous loss of angular momentum of a spinning cosmic string due to 
particle emission is discussed. The rate of particle production between 
the spinning cosmic string spacetime ($t \rightarrow -\infty $)
and a non-spinning cosmic string spacetime ($t \rightarrow +\infty $) 
is calculated.

\end{abstract}

\end{center}

\vspace{1cm}


\vfill{\footnotesize \tt \mbox{}\\
\line(1,0){180}
\par $\mbox{}^{*}$ lorenci@cpd.efei.br
\par $\mbox{}^{\dag}$ rpaola@lafex.cbpf.br
\par $\mbox{}^{\ddag}$ nfuxsvai@lafex.cbpf.br}

\newpage

\section{Introduction}

One important property of field theories with spontaneous symmetry
breaking is that the life-time of energetically meta-stable vacuum
states can be very long. This can lead to the partitioning of the
universe into regions of different meta-stable vacuum states. One
of these structures is called cosmic strings \cite{Kibble}. 
Such kind of structures represent thin
tubes of false vacuum and are expected to be of large linear mass density
$\mu$ and can also carry intrinsic angular momentum $J$.

The cosmic strings affect the spacetime mainly topologically, giving a 
conical structure to the space region around the cone of the string. 
The conical topology  may be responsible for several gravitational effects, 
gravitational lensing \cite{Vilenkin} and particle
production due to the changing gravitational field during the formation 
of such object \cite{Parker} being some examples.

There is a lot of papers studying quantum processes in a cosmic 
string spacetime. Of special interest to us are the following: \cite{Harari} 
where pair production in a straight cosmic string spacetime is discussed, 
\cite{Nami} where the response function of detectors in the presence of
the cosmic strings is calculated, and \cite{Mazur} where the spinning 
cosmic string spacetime is investigated. Finally spinning cosmic strings
were studied also by Gal'tsov {\it et al.} and also Letelier. These authors 
investigated chiral strings generating a chiral conical spacetime 
\cite{galtsov}.

In this paper we are interested in two calculations. First, to analyse the
energy momentum distribution of a time dependente spinning cosmic string
spacetime. Second, to estimate the particle production due
to the changing gravitational field in the situation of gradual loss
of angular momentum.  

\section{General spinning cosmic string spacetimes}

The Einstein\rq s gravity equation of general relativity is given
by:
\begin{equation}
G_{\mu\nu} = -\kappa T_{\mu\nu}
\label{1}
\end{equation}
where the constant $\kappa$ is defined to give the correct Newtonian 
limit, resulting in
\begin{equation}
\kappa = \frac{8\pi G}{c^4}.
\label{2}
\end{equation}
$G$ and $c$ appearing in the latter equation represent the Newtonian 
gravitational constant and light velocity, respectively.

In order to obtain the geometry generated by a rotating cosmic string
we proceed in the following way. First, we choose a cylindrical coordinate
system ($x^{\mu}:\{ct,r,\varphi,z\}$) in which an infinitely long and
thin cosmic string lies along the $z$-axis. We consider a stationary
string carrying linear densities of mass $\mu$ and angular momentum
${\cal J}_{\scriptscriptstyle 0}$. The mass density is proportional to a 
two-dimensional
delta function ($\delta$-function) while the angular momentum density
is proportional to the derivatives of $\delta$-functions. The
$T_{\scriptscriptstyle 00}$ and $T_{\scriptscriptstyle 0i}$ 
components of the associated energy momentum tensor
will be proportional to $\delta$-function and its derivatives, respectively%
\footnote{A good discussion on this topic can be found in the papers of
Gal'tsov and Letelier \cite{galtsov}.}. Thus, Einstein equations leads
to the geometry  
\begin{equation}
[g_{\mu\nu}] = 
\left( 
     \begin{array}{cccc}
      c^2 &  0 & 4G{\cal J}_{\scriptscriptstyle 0}/c^2  & 0 \\
      0 & -1 & 0 & 0 \\
      4G{\cal J}_{\scriptscriptstyle 0}/c^2 & 0  & 
-\left(b^2r^2-16G^2{\cal J}_{\scriptscriptstyle 0}^2/c^6\right) & 0 \\
      0 &           0           &           0        & -1
     \end{array}
\right)
\label{3}
\end{equation}
with 
\begin{equation}
b \equiv 1-\frac{4G\mu}{c^2}.
\label{4}
\end{equation}
From these metric components we can construct the associated line element
\begin{equation}
ds^{2} = 
\left( cdt + \frac{4G{\cal J}_{\scriptscriptstyle 0}}{c^3}
d\varphi\right)^{2} - dr^{2}
- b^{2}r^{2}d\varphi^{2}-dz^{2}.
\label{5}
\end {equation}
Therefore, as was noticed by Deser and Jackiw \cite{Deser92}, the two-parameter
$(\mu;{\cal J}_{\scriptscriptstyle 0})$ metric tensor showed in equation (\ref{3})
represents the general time-independent solution to gravitational Einstein
equations outside any matter distribution lying in a bounded region on the
plane and having cylindrical symmetry.
 In this work we will consider only the
exterior region of the cosmic string. Thus the $\delta$-functions will
be suppressed in our analysis of the energy momentum tensor  generating
a spacetime configuration due to a cosmic string loosing angular momentum.

In the same way the exact spacetime metric representing the vacuum solution
of a static cylindrically symmetric cosmic string is found \cite{Hisc85} to
be
\begin{equation}
ds^{2} = c^2dt^2 - dr^{2}
- b^{2}r^{2}d\varphi^{2}-dz^{2}.
\label{6}
\end {equation}
Obviously this metric is a particular case of the former, equation (\ref{5}), 
with null density of angular momentum. 

Thus it arises the questions: How can we go from a spinning cosmic string
spacetime to a static one and what is the rate of particle production
between these two asymptotic limits?

First of all, let us assume that the cosmic string is generated in such way
that there is spontaneous loss of energy associated with change in its 
rotation velocity, i.e., loss of density of angular momentum. Quantitatively
such process would correspond to a non-stationary spacetime geometry that
could be described by the metric (\ref{5}) with 
${\cal J}_{\scriptscriptstyle 0}$ replaced by a general function of time
${\cal J}(t)$:
\begin{equation}
ds^{2} = 
\left[ cdt + \frac{4G{\cal J}(t)}{c^3}
d\varphi\right]^{2} - dr^{2}
- b^{2}r^{2}d\varphi^{2}-dz^{2}
\label{7}
\end {equation}
and provided with the asymptotic conditions
\begin{eqnarray}
\lim_{t\rightarrow -\infty}{\cal J}(t) &=& {\cal J}_{\scriptscriptstyle 0}
\label{8} \\
\lim_{t\rightarrow +\infty}{\cal J}(t) &=& 0.
\label{9}
\end{eqnarray}
A specific choice for the density of angular momentum that 
solves the above relations is given by
\begin{equation}
{\cal J}(t) = \frac{{\cal J}_{\scriptscriptstyle 0}}{2}
\left[1-\tanh(t)\right].
\label{10}
\end{equation}

In the following section we will look for the kind of matter
content that generates such spacetime geometry.

\section{Energy momentum distribution}

As we can notice, the region between the two asymptotic limits represents
a curved spacetime. From Einstein equations we show that the fluid 
configuration characterizing such geometry is represented by the following
energy momentum tensor components
\begin{equation}
[T_{\mu\nu}] = 
-\frac{2G{\cal J}_{\scriptscriptstyle 0}}{\kappa r \cosh t}\left( 
     \begin{array}{cccc}
      0 & 0 & 0 & 0 \\
      0 & \frac{2GJ_{\scriptscriptstyle 0}}{b^2 r}F(t) & 1 & 0 \\
      0 & 1 &  0  & 0 \\
      0 & 0 &  0  & \frac{2GJ_{\scriptscriptstyle 0}}{b^2 r}F(t)
     \end{array}
\right)
\label{Energia}
\end{equation}
with $F(t)$ defined by:
\begin{equation}
F(t) \equiv \frac{3+\sinh(2t)-2\cosh^2(t)}{\cosh^3(t)}.
\end{equation}

Let us now analyze the properties of the above characterized fluid. 
First of all we define a 4-velocity vector field $V^{\mu}$
\begin{equation}
V^{\mu} = \delta^{\mu}_{0}
\end{equation}
and the projector tensor $h_{\alpha\beta}$
\begin{equation}
h_{\alpha\beta} = g_{\alpha\beta} - V_{\alpha}V_{\beta}.
\end{equation}
In the standard way, we express the energy momentum tensor in terms of its
irreducible parts as
\begin{equation}
T_{\mu\nu} = (\rho + p)V_{\mu}V_{\nu} - p g_{\mu\nu} + q_{(\mu}V_{\nu)}
+ \Pi_{\mu\nu}
\end{equation}
where we introduced the quantities related with the fluid, i.e., energy
density, isotropic pressure, heat flux and anisotropic pressure, defined 
respectively as
\begin{eqnarray}
\rho & = & T_{\mu\nu} V^{\mu} V^{\nu} \\
p & = & -\frac{1}{3} T_{\mu\nu} h^{\mu\nu} \\
q_{\alpha} & = & T_{\mu\nu} V^{\nu} h^{\mu}\mbox{}_{\alpha} \\
\Pi_{\mu\nu} & = & T_{\alpha\beta}h^{\alpha}\mbox{}_{\mu}
h^{\beta}\mbox{}_{\nu} + p h_{\mu\nu}.
\end{eqnarray}
For the case we are analyzing here only isotropic and anisotropic
pressure survive, and they result in:
\begin{equation}
p(t,r) = -\frac{8 G^2 J_{\scriptscriptstyle 0}^2}{3 \kappa b^2 c^8 r^2}
\left( \tanh^2t - 1\right)\left(3\tanh^2t - 2\tanh t -1 \right)
\end{equation}
and
\begin{equation}
[\Pi_{\mu\nu}] = 
-\frac{p(t,r)}{2} \left( 
     \begin{array}{cccc}
      0 & 0 & 0 & 0 \\
      0 & 1 & \tilde{\pi} & 0 \\
      0 & \tilde{\pi} &-2 & 0 \\
      0 & 0 & 0 & 1
     \end{array}
\right)
\label{energia}
\end{equation}
where
\begin{equation}
\tilde{\pi} \equiv -\frac{3 b^2 c^4 r}{2 G J_{\scriptscriptstyle 0}}
\left(\frac{1}{3\tanh^2t - 2\tanh t -1}\right).
\end{equation}
From what we have seen, outside the core of the string there is not
energy density of matter but only flux of
energy, that appears as pressure. From the Einstein equations we obtain the 
relation between the trace of the energy momentum tensor and the curvature
of the spacetime ($R = \kappa T$), that results in:
\begin{equation}
R = -3 \kappa p(t,r).
\end{equation}
\begin{figure}[htb]
\leavevmode
\centering
\epsfxsize=80ex
\epsfysize=80ex
\epsfbox{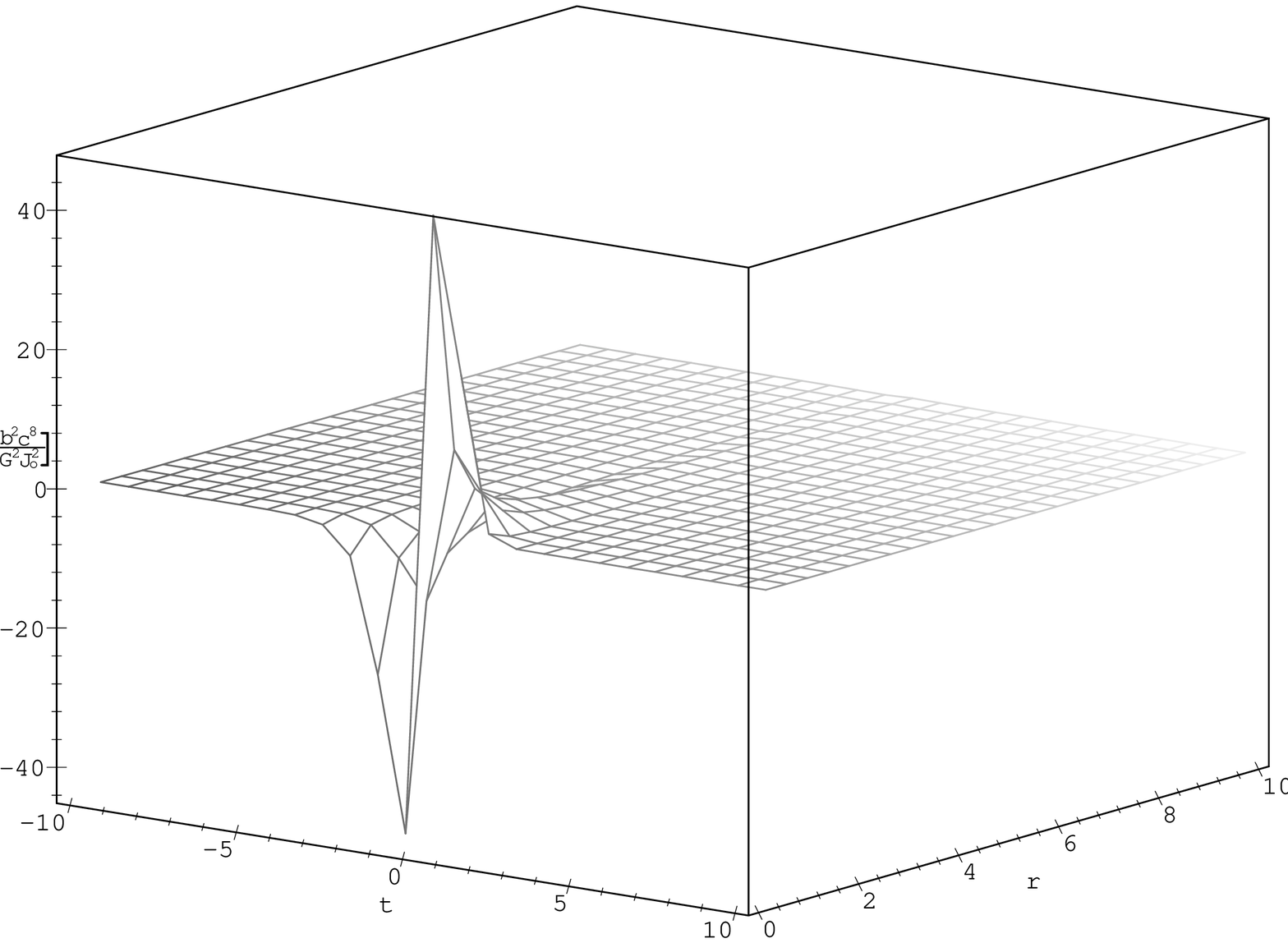}
\protect\vspace{-7\baselineskip}
\caption{The spacetime curvature $R$ (as function of coordinates $t$ and $r$)
generated by a spinning cosmic string loosing angular momentum during the 
time evolution.}
\label{figure}
\end{figure}
The geometry of the spacetime is flat in the asymptotic regions
corresponding to infinite past and infinite future. Therefore it is worth
to analyze the behavior of the curvature in the intermediate region. In the
figure \ref{figure} we plot the curvature $R$ as function of time $t$ and
radial coordinate $r$ (distance from the $z$ axes of the string). For $t << 0$
the spacetime is approximately flat for all values of $r$, and for large 
values of $r$ it will still be flat for all values of $t$, as we would
expect. As time
$t$ goes from negative to positive values the curvature
decreases until it reaches a minimum value at $t \approx -0.926$. 
After this point it begins to increase, becomes positive and reaches a 
maximum value at $t \approx 0.233$.
Finally it asymptotically goes to zero. The two asymptotic spacetimes are
flat. The only difference between them is that for 
$t \rightarrow -\infty $ the string is spinning with constant density of
angular momentum $J_{\scriptscriptstyle 0}$.

As a general result of this analysis we obtained that the loss of angular
momentum by a cosmic string causes a change in the metric properties of
the spacetime. But changes in the gravitational field generate particles 
\cite{Birrell}. Thus we have to investigate the creation of particles and 
radiation by the changing gravitational field during the loss of angular 
momentum of a initially spinning cosmic string. We will treat this problem 
by performing the calculation of the Bogoliubov transformation entailed by 
this process.

\section{Particle production between the asymptotic spacetimes}

From the general geometry (\ref{7}) and the conditions (\ref{8}) and
(\ref{9}), in the two asymptotic regions --- the infinite past and 
infinite future --- the spacetime metric structures reduce to:
\begin{eqnarray}
ds^{2}_{-\infty} &=& 
\left( dt + 4GJ_{\scriptscriptstyle 0} 
d\varphi\right)^{2}-dr^{2}
- b^{2}r^{2}d\varphi^{2}-dz^{2}
\label{V6}\\
ds^{2}_{+\infty} &=& 
dt^{2}-dr^{2}-b^{2}r^{2}d\varphi^{2}-dz^{2}.
\label{V7}
\end{eqnarray}
The above metrics represent the structure of the
spacetime in the exterior region of a rotating and a non-rotating
cosmic string, respectively. 

In this section we will analyze the rate of particle production by the
changing gravitational field between the two asymptotic spacetimes
during the evolution of a cosmic string that looses angular momentum.
A similar idea was used by Bernard and Duncan \cite{Duncan} to study
a two dimensional Robertson-Walker model where the
conformal scale factor has the same functional form as equation (\ref{10}). 
In the two asymptotic limits, the spacetime becomes Minkowskian, and 
they obtained the mode solutions of the Klein-Gordon
equation in these two limits. A straightforward calculation of the Bogoliubov 
coefficients between the {\em in} and {\em out} modes gives the rate of
particle production during the expansion of the universe. It is worth to 
notice that such method was first introduced by L. Parker \cite{ParkerT}
in order to calculate particle creation by the expansion of the universe.

Let us consider a massive minimally coupled Hermitian scalar 
field $\phi(t,r,\varphi,z)$ defined at all points of the 
spacetime (\ref{7}). The Klein-Gordon equation is given by:
\begin{equation}
\left[g^{\mu\nu}D_{\mu}D_{\nu} + M^{2}\right]\phi(t,r,\varphi,z) = 0
\label{V8}
\end{equation}
where the symbol $D_{\alpha}$ represents the covariant derivative 
with respect to the metric $g_{\alpha\beta}$, and $M$ is the mass
of the scalar field. 

To maintain the energy produced in a limited region of the
space we impose Dirichlet boundary conditions at $r=R$,
\begin{equation}
\left.\phi(t,r,\varphi,z)\right|_{r=R} = 0
\label{R}
\end{equation}
and periodic boundary conditions in $z$ with period $L$.

In order to circumvent the problem of the generation of closed time-like
curves (CTC), we impose an additional
vanishing boundary condition at $r=R_{0} > 4GJ/b$. Thus, the radial coordinate
has the domain $R_{0} < r < R$. For a careful study on how to construct
quantum field theory in a spacetime with CTC, see for instance
\cite{Boul}. The same problem appears in $(2+1)$ dimensional gravity since
the spinning cosmic string spacetime is exactly the solution of $(2+1)$ 
Einstein equations of a spinning point source \cite{Deser}.

The Klein-Gordon equation for the initial spacetime (\ref{V6}) reduces
to the form:
\begin{equation}
\left[\left(1-\frac{16 G^2 J_{\scriptscriptstyle 0}^{2}}
{b^{2}r^{2}}\right)\frac{\partial^2}{\partial t^2}
-\frac{1}{r}\frac{\partial}{\partial r}
\left(r\frac{\partial}{\partial r}\right)
-\frac{1}{b^{2}r^{2}}\frac{\partial^{2}}{\partial\varphi^{2}}
-\frac{\partial^{2}}{\partial z^{2}}
+ \frac{8 G J^{2}_{\scriptscriptstyle 0}}
{b^2r^2}\frac{\partial^2}{\partial t \partial \varphi}  
+M^{2}\right]\phi(t,r,\varphi,z)=0. 
\label{V9}
\end{equation}
The mode solutions $u_{j}$ are found to be%
\footnote{We are using a collective index $j=\{l,m,n\}$.}
\begin{equation}
u_{j}(t,\vec{x}) = N_{1}\,e^{-i\omega_{l}t}e^{ikz}e^{im\varphi}
\left[\frac{J_{\mu}(qr)}{J_{\mu}(qR)} - \frac{Y_{\mu}(qr)}{Y_{\mu}(qR)}
\right]
\label{V16}
\end{equation}
with
\begin{eqnarray}
\mu &\equiv& \frac{|m+4GJ\omega_{l}|}{b}\\
q &=& \sqrt{\omega_{l}^{2} - k^{2} - M^{2}}
\label{V34}
\end{eqnarray}
and 
\begin{equation}
k = \frac{2\pi n}{L}.
\end{equation}
In the above, $m,n=0,\pm 1,\pm 2,...$. Choosing the constant $N_{1}$ to 
make the set orthonormal results:
\begin{equation}
N_{1} = (2\omega_{l})^{-\frac{1}{2}}\left\{ 
V \left[\frac{J^{'}_{\mu}(qR)}{J_{\mu}(qR)} - 
\frac{Y^{'}_{\mu}(qR)}{Y_{\mu}(qR)} \right]^2
-  V_{0}\left[\frac{J^{'}_{\mu}(qR_{0})}{J_{\mu}(qR)} - 
\frac{Y^{'}_{\mu}(qR_{0})}{Y_{\mu}(qR)} \right]^2
\right\}^{-\frac{1}{2}}
\label{v33}
\end{equation}
where we defined the 3-volumes $V \equiv b\pi LR^2$ and $V_{0} \equiv 
b\pi LR_{0}^2$. The values of the parameter $q$ are determined by 
a transcendental equation  which comes from the vanishing boundary conditions,
that is,
\begin{equation}
J_{\mu}(qR_0)Y_{\mu}(qR) - 
J_{\mu}(qR)Y_{\mu}(qR_0) = 0
\end{equation}
and the infinite number of its roots are labeled by $l=1,2,3,...$.

The modes $u_{j}(t,\vec{x})$, form a basis in the space of the solutions of 
the Klein-Gordon equation and can be used to expand the field operator as:
\begin{equation}
\phi_{in}(\vec{x},t) = \sum_{j}\left[ a_{j} u_{j}(t,\vec{x}) 
+ a_{j}^{\dag}u_{j}^{*}(t,\vec{x})\right].
\label{V35}
\end{equation}

The creation and annihilation operators $a^{\dag}_{j}$ and $a_{j}$ 
satisfy the commutation relation:
\begin{equation}
[ a_{j},a^{\dag}_{j'}] = \delta_{j,j'}
\label{aa}
\end{equation}
and we define the  in-vacuum state by
\begin{equation}
a_{j}\left|0,in\right> = 0 \,\,\,\, \forall\,\, j.
\label{0in}
\end{equation}

In the same way we can perform the canonical quantization of the field in the 
infinite future. The Klein-Gordon equation in the non-rotating cosmic string 
spacetime (\ref{V7}) reads
\begin{equation}
\left[\frac{\partial^2}{\partial t^2} -
\frac{1}{r}\frac{\partial}{\partial r}\left(r\frac{\partial}{\partial r}\right)
-\frac{1}{b^{2}r^{2}}\frac{\partial^{2}}{\partial\varphi^{2}}-
\frac{\partial^{2}}{\partial z^{2}}
+M^{2}\right]\phi(t,r,\varphi,z)=0.
\label{V13}
\end{equation}
The mode solutions $v_{j}$ are found to be:
\begin{equation}
v_{j}(t,\vec{x}) = N_{2}\,e^{-i\Omega_{l'}t}e^{ik'z}e^{im'\varphi}
\left[\frac{J_{\nu}(\bar{q}r)}{J_{\nu}(\bar{q}R)} - 
\frac{Y_{\nu}(\bar{q}r)}{Y_{\nu}(\bar{q}R)}\right]
\label{V37}
\end{equation}
with
\begin{eqnarray}
\nu &\equiv& \frac{|m'|}{b}\\
\bar{q} &=& \sqrt{\Omega_{l'}^{2} - k'^{2} - M^{2}}
\label{nu}
\end{eqnarray}
and 
\begin{equation}
k' = \frac{2\pi n'}{L}.
\end{equation}
Choosing the constant $N_{2}$ in order to make the set of modes 
$\{v_{j},v^{*}_{j}\}$ orthonormal, results:
\begin{equation}
N_{2} = (2\Omega_{l'})^{-\frac{1}{2}}
\left\{ 
V\left[\frac{J^{'}_{\nu}(\bar{q}R)}{J_{\nu}(\bar{q}R)} - 
\frac{Y^{'}_{\nu}(\bar{q}R)}{Y_{\nu}(\bar{q}R)}\right]
-  V_{0}\left[\frac{J^{'}_{\nu}(\bar{q}R_{0})}{J_{\nu}(\bar{q}R)} - 
\frac{Y^{'}_{\nu}(\bar{q}R_{0})}{Y_{\nu}(\bar{q}R)}\right]
\right\}^{-\frac{1}{2}}
\label{V33}
\end{equation}
where the values of $\bar{q}$ are determined by a transcendental equation
of the same type as before.

The out modes, solutions of the Klein-Gordon equation, also form a 
complete set and can be used to expand the field operator as: 
\begin{equation}
\phi_{out}(t,\vec{x}) = \sum_{j}\left[ b_{j} v_{j}(t,\vec{x}) 
+ b_{j}^{\dag}v_{j}^{*}(t,\vec{x})\right].
\label{V36}
\end{equation}
The creation and annihilation operators $b^{\dag}_{j}$ and 
$b_{j}$ satisfy the usual commutation relation:
\begin{equation}
[ b_{j},b^{\dag}_{j'}] = \delta_{j,j'}
\label{bb}
\end{equation}
and the out-vacuum state is defined by
\begin{equation}
b_{j}\left|0,out\right> = 0 \,\,\,\, \forall \,\, j.
\label{0out}
\end{equation}

Following Parker we will calculate the rate of particle production between 
two asymptotic spacetimes discussed above: the spinning 
cosmic string spacetime in the infinite past and a non-spinning cosmic string 
spacetime in the infinite future. 

An important point is that in our model we have not to deal with
the problems of junction conditions since there is no sudden approximation
here. The metric evolves continuously between both asymptotic states. The
angular momentum of the spinning cosmic string is lost by 
particle emission processes. The fundamental quantity we have to calculate
is the Bogoliubov coefficients between the modes in the non-spinning and
spinning cosmic string spacetime. The average number of out-particles in the modes $j=(l,m,n)$ produced by this process is:
\begin{equation}
\left<in,0\right| b^{\dag}_{j}b_{j} \left|0,in\right> = 
\sum_{i}\left|\beta_{ij}\right|^{2}.
\label{number}
\end{equation}
Using the definition of the Bogoliubov coefficients $\beta_{ij}$ given by
\begin{equation}
\beta_{jj'} = -(u_{j},v^{*}_{j'})
\end{equation}
we have 
\begin{equation}
\beta_{jj'} = -2\pi b L(\Omega_{l'} -
 \omega_{l})N_{1}N_{2}\xi(R,R_{0})\delta_{m,-m'}\delta_{n,-n'}
\label{beta}
\end{equation}
where
\begin{equation}
\xi(R,R_{0}) \equiv \int^{R}_{R_{0}} dr\, r\,
\left[\frac{J_{\mu}(qr)}{J_{\mu}(qR)} - \frac{Y_{\mu}(qr)}{Y_{\mu}(qR)}
\right]\left[\frac{J_{\nu}(\bar{q}r)}{J_{\nu}(\bar{q}R)} - 
\frac{Y_{\nu}(\bar{q}r)}{Y_{\nu}(\bar{q}R)}\right]
\end{equation}

By substituting (\ref{beta}) in equation (\ref{number}) and using the definitions 
of the normalization constants $N_1$ and $N_2$, the average number  of 
particles produced in the mode $j = (l,m,n)$ is:

\begin{eqnarray}
\left<in,0\right| b^{\dag}_{j}b_{j} \left|0,in\right> 
&=& \frac{1}{4} \left[\left(\frac{\Omega_{l'}}{\omega_{l}}\right)^{\frac{1}{2}}
- \left(\frac{\omega_{l}}{\Omega_{l'}}\right)^{\frac{1}{2}}\right]^2 
\nonumber\\
&\times&\hspace{-3mm} \left\{ 
V \left[\frac{J^{'}_{\mu}(qR)}{J_{\mu}(qR)} - 
\frac{Y^{'}_{\mu}(qR)}{Y_{\mu}(qR)} \right]^2
-  V_{0}\left[\frac{J^{'}_{\mu}(qR_{0})}{J_{\mu}(qR)} - 
\frac{Y^{'}_{\mu}(qR_{0})}{Y_{\mu}(qR)} \right]^2
\right\}^{-1}.
\end{eqnarray}

The expression of the average number of particles in the mode $j$ is very
complicated, and some simplifications used by Mendell and Hiscock can not 
be made here. The key point is that in Parker's work use was made of the
sudden approximation, i.e., for $t-\epsilon$ the spacetime is Minkowskian
and for $t+\epsilon$ the geometry is conic. Mendell and Hiscock extended
Parker's work considering a number of different models still using the
sudden approximation. Consequently if $\Delta t$ is the actual time of 
formation of the string, the production of particles in modes with frequencies
that are large compared with $\Delta t^{-1}$ are suppressed. In our model,
the process of particle production by loss of angular momentum takes an 
infinite time, since we have two asymptotically flat spacetimes with curved
geometry between both states. Nevertheless some conclusions can 
be obtained. Note that the part of the above expression that comes between
braces does not depend on the energy of the produced particle $\Omega_{l'}$.
From this it is clear that the number of produced particles diverges for
low and high energy modes. This behavior is expected for the low energy
modes, but is not for the high ones. A way to improve our model is to
assume a finite time $\Delta t_0$ for the loss of the angular momentum to 
occur, thereby introducing a natural cutoff in the energy of the produced particles, because now the production of modes for which the energy is large compared with $\Delta t_0^{-1}$ will be also suppressed, as in the case of the
sudden approximation.

\section{Conclusion}

In this paper we discussed the properties of a fluid generating 
a general spinning
cosmic string spacetime with flat limiting cases corresponding to
a constant angular momentum in the infinite past and static 
configuration in the infinite future. We also analyze the  
particle production by loss of angular momentum in a spinning cosmic string
spacetime. To circumvent the problem of CTC\rq s we assumed a cosmic string 
with a radius fixed. Moreover, to keep energy and number of particles
produced by the process in a finite region of space,
following Parker's arguments, we impose vanishing 
boundary conditions on the wall of a cylinder with 
finite radius $R$.

A possible continuation of this paper is to formulate the energy 
conservation law, that is, to show if there is a balance between the 
total energy of the particles created and the energy associated 
with loss of angular momentum. This can be done comparing the vacuum 
stress-tensor of the massive field in the spinning and non-spinning 
cosmic string spacetimes. The calculation for a massless conformally coupled 
scalar field in the non-spinning cosmic string spacetime
has been done by many authors \cite{Linet}. The same calculation in 
the spinning cosmic string spacetime has been done by \cite{Matsas}. 
As far as we know the renormalized stress tensor of a massive minimally 
coupled scalar field  has not been investigated in the literature.

\section{Acknowledgement}

This paper was partially supported by {\it Conselho Nacional de Desenvolvimento 
Cient\'{\i}fico e Tecnol\'ogico} (CNPq), of Brazil.

\end{document}